\documentclass[aps,showpacs,pre,twocolumn,reprint]{revtex4-1}

\usepackage{graphicx,bm,amsmath,amssymb,natbib,url,epsfig}
\usepackage{dcolumn} 
\usepackage{hyperref}
\usepackage{verbatim} 
\usepackage[outdir=./]{epstopdf}

\usepackage{color}

\definecolor{darkgreen}{rgb}{0.1,.6,.1}
\definecolor{greenblue}{rgb}{0.0,.1,.4}

\usepackage[normalem]{ulem}

\allowdisplaybreaks

\begin{document}

\date{}



      

  
  
  






\title{\Large Role of limiting dispersal on metacommunity stability and persistence}


  \bigskip
  
  \bigskip
  
  \author{Snehasish Roy Chowdhury${}^{1}$}  
  
\author{Ramesh Arumugam${}^{1}$}  
 
 \author{Wei Zou${}^{2}$}

\author{ V K Chandrasekar${}^{3}$}

\author{D V Senthilkumar${}^{1}$}

\affiliation{${}^{1}$School of Physics, Indian Institute of Science
  Education and Research, Thiruvananthapuram - 695551, Kerala, India
\\{$^2$ School of Mathematical Sciences\text{,} South China Normal University\text{,} Guangzhou\text{,} 510631\text{,} P. R. China}
  \\ ${}^{3}$Department of Physics, Centre for Nonlinear Science and Engineering,~School of
  Electrical and Electronics Engineering, SASTRA Deemed University,
  Thanjavur - 613401, Tamilnadu, India }


\date{\today}

\begin{abstract}

The role of dispersal on the stability and synchrony of a metacommunity is a topic of considerable interest in theoretical ecology. 
Dispersal is known to promote both synchrony, which enhances the likelihood of extinction, and spatial heterogeneity, which favours the persistence of the population. 
Several efforts have been made to understand the effect of diverse variants of dispersal in the spatially distributed ecological  community. 
Despite the environmental change strongly affect the dispersal,
 the effects of controlled dispersal on the  metacommunity stability and their persistence  remain unknown.  
 We study the influence of limiting the immigration using two patch  prey-predator metacommunity at both local and spatial scales. 
 We  find that the spread of the inhomogeneous stable
steady states (asynchronous states)   decreases monotonically  upon limiting the predator dispersal.
 Nevertheless, at the local scale,  the spread of the inhomogeneous steady states increases  up to a critical value of the limiting factor,  
favoring the metacommunity persistence,  and then starts decreasing  for further decrease in the limiting factor with varying local interaction.  
Interestingly, limiting the prey dispersal promotes inhomogeneous steady states in a 
large region of the parameter space, thereby increasing the metacommunity persistence both at spatial and local scales.
Further, we show similar qualitative dynamics in an entire class of complex networks consisting of  a large number of patches.
 We  also deduce various bifurcation
curves and stability condition for the  inhomogeneous steady states, which we find to agree well with the simulation results. Thus, our
findings on the effect of the limiting dispersal can help to develop conservation measures for ecological community.

\end{abstract}

\maketitle
\bigskip

\section{Introduction}

\bigskip

Synchrony, referring the coherent behaviour of coupled systems \cite{PiRo01}, has been  studied extensively in numerous fields such as neuroscience, physiology, biology, ecology \cite{Izh07,BlHu99,LiKo04}, etc.  In population dynamics, synchrony has received a great deal of attention  since it can elevate a high risk of extinction \cite{RaKa95,LiKo04,BrHo04,GoHa08}. Indeed, understanding the factors and  mechanisms that generate population synchrony in  ecology is of great importance for conservation and ecosystem management. Three major mechanisms of population synchrony are dispersal, environmental variation and trophic interactions \cite{RaKa97,KeBj00,LiKo04}. Among these, dispersal is a widely studied phenomenon in population dynamics \cite{HeKa97,BlHu99,GoHa08,HoHa08}. Interestingly, dispersal not only  leads to synchronized behaviour among the spatially distributed populations, but it can also rescue populations through recolonization \cite{Doe95,Han98,Han99}. Consequently, dispersal can also act as a stabilizing  mechanism in diverse systems and has a huge impact on the persistence of ecological communities \cite{BrHo04,WaLo16}. In this scenario, a small change or a control over the dispersal can have tremendous consequences on the stability and ecosystem functioning. In this study, we present how limiting the  dispersal  of both predator and prey populations can affect  the stability and persistence of a spatially distributed community. 

\bigskip
\begin{figure*}[tbh]
 \centering
\includegraphics[width=0.85\textwidth,angle=0]{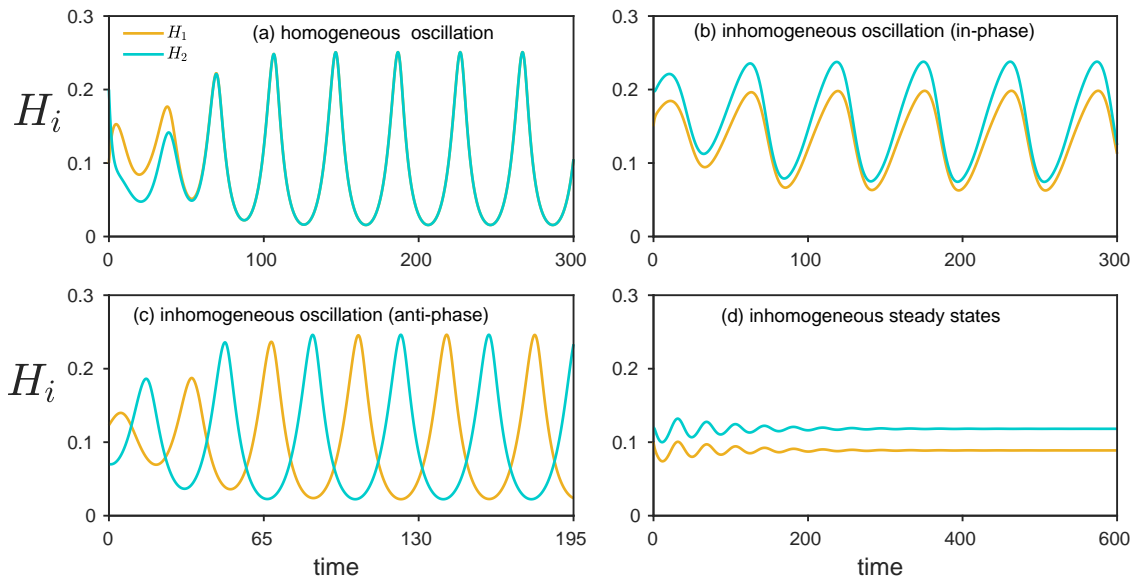} 
\caption{Homogeneous and inhomogeneous stable states of the metacommunity model: (a) homogeneous oscillatory state for $d_h=0.2$ and $k=0.2$, (b) inhomogeneous oscillatory states are in in-phase for $d_h=1$ and $k=1.2$, (c)  inhomogeneous stable oscillatory states are in anti-phase for $d_h=0.001$ and $k=0.6$, and (d) inhomogeneous steady states for $d_h=0.45$ and $k=0.6$. Other parameters are $r=0.5$, $a=1$, $c=0.5$, $b=0.16$, $m=0.2$, $d_v=0.001$ and $\alpha=\beta=1$.}
  \label{fig1}
\end{figure*}

Climate change affects the stability of  many ecological communities by altering their  dispersal pattern and the associated responses \cite{McHo92,DoRu97,TrDe13,FuSc14}. Major ecological  responses to environmental changes include  adaptation, migration and extinction \cite{Wa02,Par06,BeKi10,Wal10,ReSc11}. Species  that lacks the adaptation for environmental variation tends to migration to suitable habitats. While species migrate, it can face an additional mortality due to  failed migration, misdirected migration and also overcrowding \cite{MuOh13}. Typically, species immigration is reduced  by habitat loss, habitat fragmentation and anthropogenic changes \cite{Wa02,Par06,PuBu11}. Indeed, climate change and other factors can also limit the species dispersal. In this  connection, theoretical studies   addressing the effect of controlled dispersal is extremely  rare. 
Despite a plethora of studies focused on understanding the role of dispersal on the metacommunity persistence, it remains unclear how a limited dispersal  affects the metacommunity stability.  Since coupled ecological oscillators constitute an efficient framework to understand the dynamical effects of dispersal, in this work, we address (i)  how a limited dispersal of both predator and prey populations affects  the stability and persistence of a spatially distributed community, and (ii) how limited dispersal along with local and spatial interactions influences the synchronized and asynchronized dynamics of the metacommunity?

\bigskip

In this study,  we address the dynamical effects of limiting the  dispersal of a predator-prey ecological system. In particular, we investigate the metacommunity persistence using homogeneous (synchronized) and inhomogeneous (asynchronized) dynamical behaviour by limiting the predator and prey dispersal. 
Our results  reveal that a small change in the predator dispersal monotonically reduces the inhomogeneous behaviour as a function of the  spatial interaction (predator/prey dispersal rates). 
In contrast, for a varying local interaction (carrying capacity), 
a small decrease in the predator  dispersal initially increases the inhomogeneous states up to a critical value of the limiting factor and eventually the inhomogeneous states  lose 
its stability resulting in homogeneous dynamical state for further decrease in the degree of the limited dispersal. Moreover, there exists a critical value of the limiting factor below which the metacommunity persists and above which metacommunity has  a risk of extinction.  On the other hand, limiting the prey dispersal manifests the stable  inhomogeneous steady
 states in a large region of the parameter space, thereby increasing the metacommunity persistence.
We also consider  an entire class of complex networks of  metacommunity to corroborate 
robustness of the results. We also deduce transcritical and Hopf bifurcation curves along with stability condition for the inhomogeneous steady state for a possible case.
We find that the stability condition agrees well with the simulation results.
 This study unveils that a controlled dispersal strongly influence the metacommunity persistence by altering the asynchronized dynamics.  

\bigskip

We organize the paper in the following order. In the `Models and methods' section, we describe a two-patch predator-prey metacommunity model with  limiting factors in the  diffusive coupling. In the `Results' section, we present the effect of limited predator and prey dispersal using  local and spatial interactions. 
We also deduce the stability condition for the inhomogeneous steady  (asynchronized alternative) states for the possible scenario. 
Subsequently, we elucidate the existence of a critical value of the limited dispersal influencing the synchronized behaviour. 
Further, we extend our analysis to the entire class of complex networks  of metacommunity to generalize our results.
Finally, we discuss  the ecological significance of our findings in the `Discussion' section.

\section{Models and methods}

\bigskip

\subsection{A metacommunity model}

We use the well-known Rosenzweig-MacArthur model~\cite{RosMac1963} to represent the  prey-predator dynamics in each patch, where
prey experiences  logistic growth and predator exhibits a Holling Type II functional response.
We consider two identical patches with diffusive coupling between them, whose governing equation is represented as
\begin{subequations}\label{eq1}
\begin{align}
\frac{dV_{i}}{dt} &= rV_{i} \left(1-\frac{V_{i}}{k} \right)-
\frac{a  V_{i}}{V_{i}+b} H_{i} +d_v(\beta V_j - V_i)\;,\\ \frac{dH_{i}}{dt} &=
 \frac{c a V_{i}}{V_{i}+b} H_{i} - m H_{i} + d_h (\alpha H_{j} -
H_{i})\;, 
\end{align}
\end{subequations}
where $i, j =1, 2$ denote the patch number and $i \neq j$.   $V_i$  and $H_i$  are the prey  and predator population densities, respectively.  
The parameter $r$ denotes the intrinsic growth rate, $k$ denotes  the carrying capacity, $a$ and $c$ denote the  maximum predation rate and predator efficiency, respectively.  
Half-saturation constant is denoted by $b$ and the mortality rate of the predator by $m$. These parameters determine the local dynamics of  the prey and the  predator in a given patch. 
The spatial interactions of the prey and predator between two patches are determined by the prey dispersal rate $d_v$, predator dispersal rate $d_h$ and the limiting factors $\alpha$ and $\beta$.
Dispersal between the patches is controlled by the limiting factors, which can be decreased from $1$ to zero. For $\alpha=\beta=1$, the coupling is the usual diffusive coupling
widely employed in the population ecology~\cite{GoHa08,HoHa08}. $\alpha<1$ accounts for the limited predator dispersal by reducing the immigration of the predator from patch $j$ to  patch $i$,
while $\beta<1$ accounts for the limited prey dispersal by reducing the immigration of the prey from patch $j$ to  patch $i$. 

\bigskip
\begin{figure*}[tbh]
 \centering
\includegraphics[width=0.7\textwidth,angle=0]{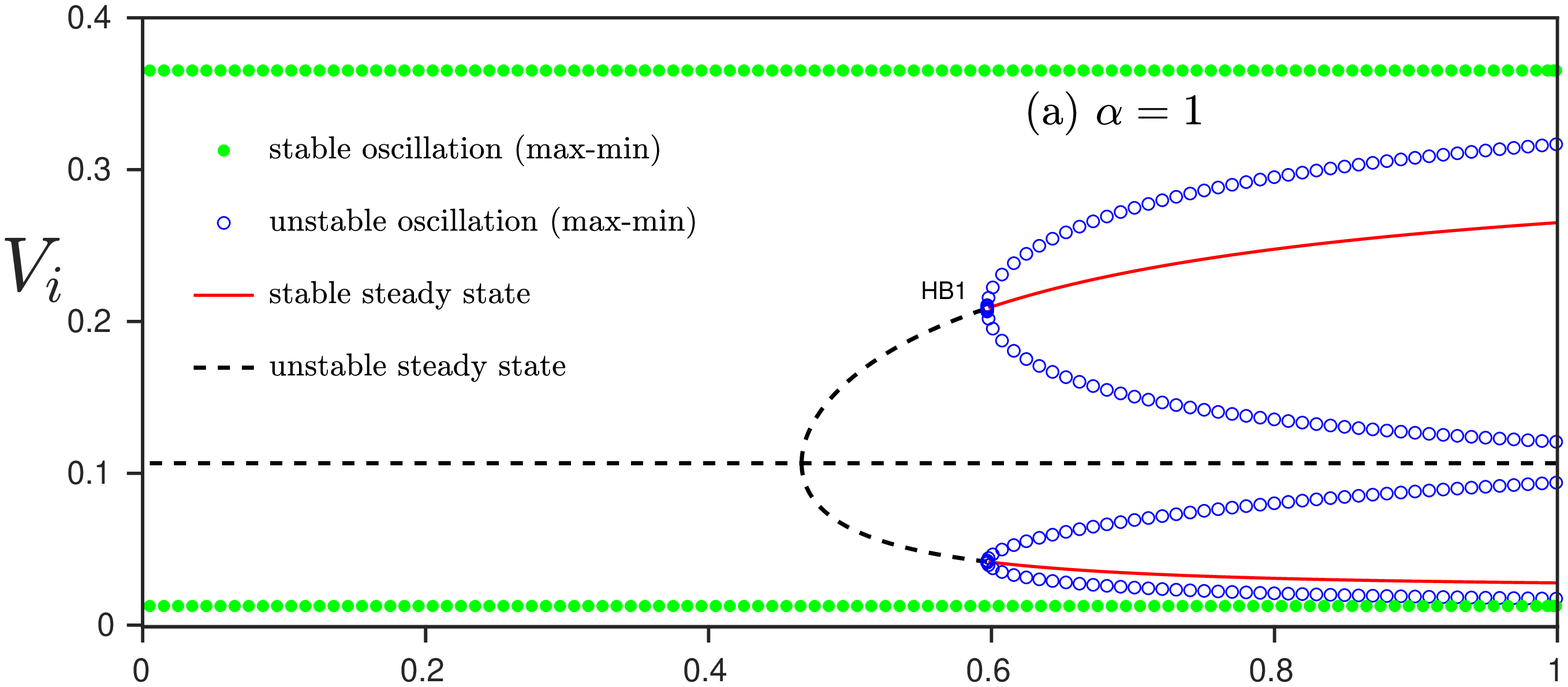}\\
\includegraphics[width=0.7\textwidth,angle=0]{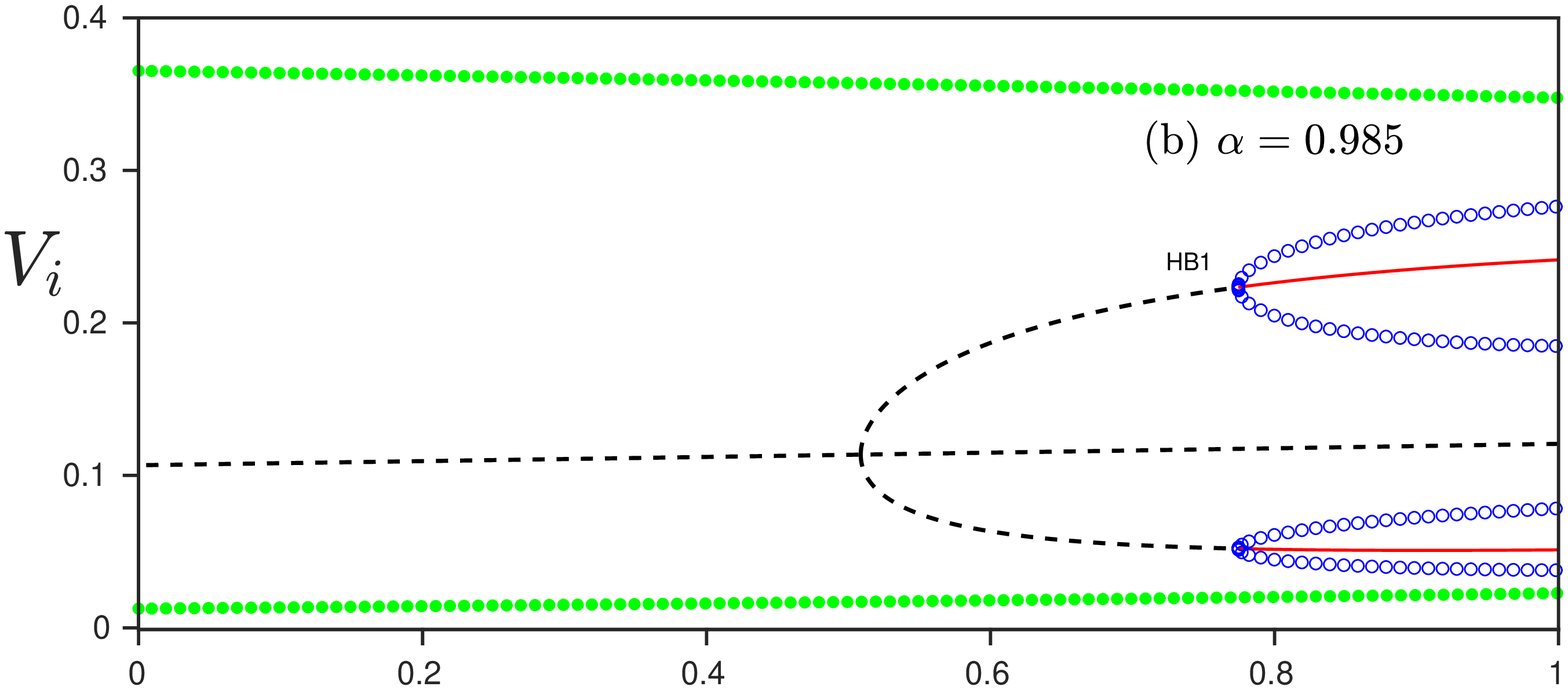}\\
\includegraphics[width=0.7\textwidth,angle=0]{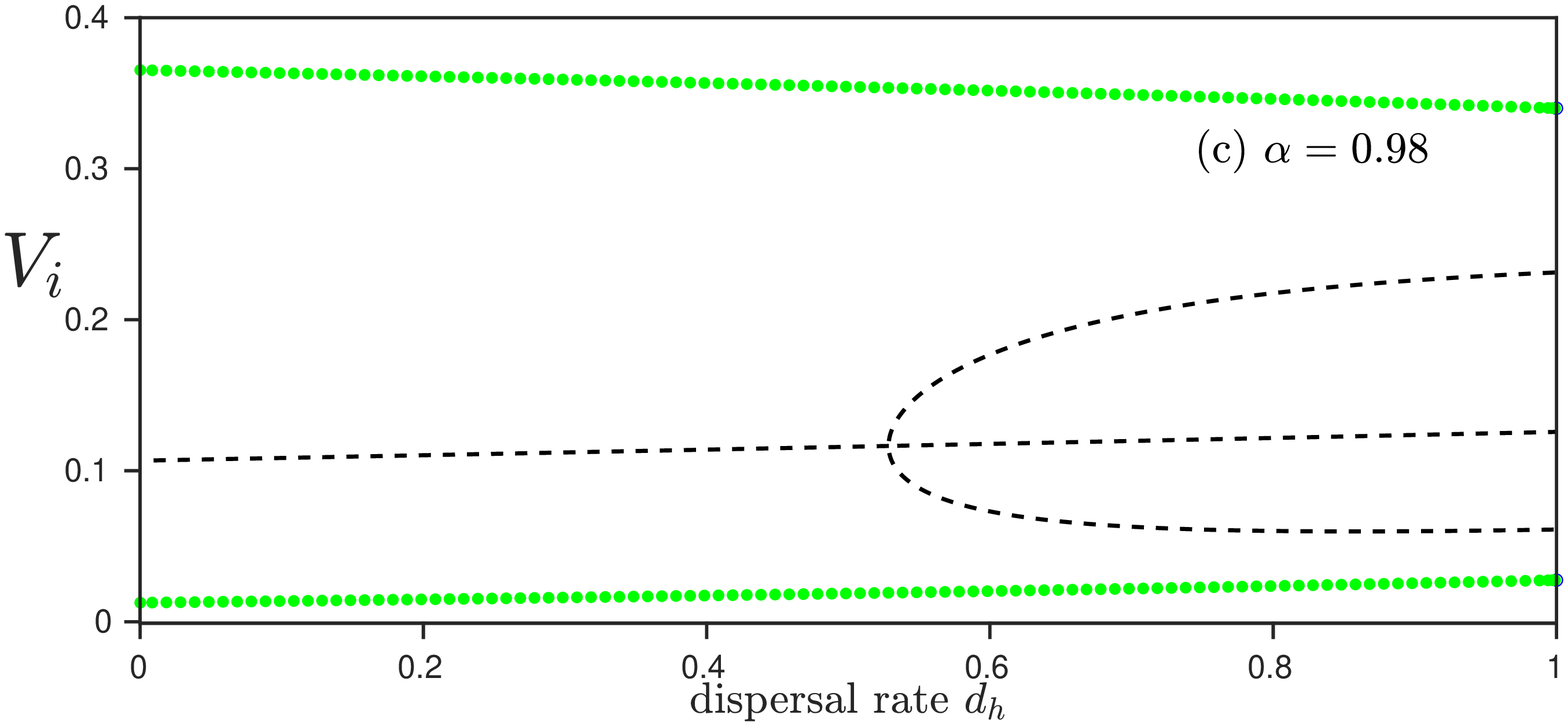}
\caption{Metacommunity dynamics for limiting the predator dispersal: Bifurcation diagram as a function of  predator dispersal rate ($d_h$) is shown for different $\alpha$ values.  Homogeneous and inhomogeneous states of the metacommunity are shown for  (a) $\alpha=1$, (b) $\alpha=0.985$, and (c) $\alpha=0.98$. For a small decrease in $\alpha$,  inhomogeneous stable steady states (red solid lines) become unstable. Here HB1 denotes  Hopf  bifurcation. Other parameters are $\beta=1$, $r=0.5$, $k=0.5$, $a=1$, $c=0.5$, $b=0.16$, $m=0.2$ and $d_v=0.001$.}
  \label{fig2}
\end{figure*}  

 Note that  the coupled
 Rosenzweig-MacArthur model with $\alpha=\beta=1$ has been widely employed to understand the intriguing collective dynamical behaviours  under various coupling 
 scenarios~\cite{psdtb2015,RaDu16,rapsd2018,ravkc2021}.   In particular, 
the  coexistence of spatially synchronized state and  death state, leading to chimeralike state,  has been reported in nonlocally coupled Rosenzweig-MacArthur model~\cite{psdtb2015},
the phenomenon of rhythmogenesis including amplitude and oscillation deaths  has been reported under the environmental coupling~\cite{RaDu16},
the effect of trade-off between the mismatch in the timescale of the species, dispersal and external force on the collective dynamics has been reported in diffusively coupled 
 Rosenzweig-MacArthur model~\cite{rapsd2018} and the effect of prey-predator dispersal on the metacommunity persistence in both constant and  temporally varying
 environment has been reported in spatially coupled Rosenzweig-MacArthur models~\cite{ravkc2021}.

\subsection{Rescaled version of coupled Rosenzweig-MacArthur model}
The above two coupled Rosenzweig-MacArthur model with $\alpha$ and $\beta$ limiting predator  and prey dispersal rates, respectively,  can also be
 written as a rescaled model with standard diffusive coupling represented as 
\begin{subequations}\label{eq1a}
\begin{align}
\frac{dV_{i}}{dt} &= r_e V_{i} \left(1-\frac{V_{i}}{k_e} \right)-
\frac{a  V_{i}}{V_{i}+b} H_{i} +\epsilon_v( V_j - V_i)\;,\\ \frac{dH_{i}}{dt} &=
 \frac{c a V_{i}}{V_{i}+b} H_{i} - m_e H_{i} + \epsilon_h ( H_{j} -
H_{i})\;, 
\end{align}
\end{subequations}
where, $r_e = r - d_v(1-\beta),  k_e = \frac{r_e}{r} k = 1 - \frac{d_v}{r}(1-\beta), \epsilon_v = d_v \beta, m_e = m + d_h(1-\alpha)$ and $\epsilon_h = d_h\alpha$.
Here, $r_e$ is the decreased growth rate, $m_e$ is the increased mortality, $k_e$ is the effective carrying capacity, and  $\epsilon_v$ and  $\epsilon_h$
are the rescaled coupling coefficients.  Since the range of $\beta$ and $\alpha$ lies within $[1,0]$,  $r_e\le r$  and $m_e\ge m$   resulting in decreased  prey growth rate and
increased predator death rate in the rescaled version of coupled Rosenzweig-MacArthur model.   Note that the results of both the rescaled model  with standard diffusive coupling and 
the original Rosenweig-MacAurther model with limiting predator and prey dispersal rates  in the coupling will be  exactly similar to each other. 
  In other words, the exact mapping between these two models can also be interpreted as that the observed results may not a pure
manifestation of the topological features.  However,
we prefer to proceed our analysis without rescaling the original Rosenweig-MacAurther model but only by engineering the coupling strategies even though
limiting the predator and prey dispersal rates effectively  results in an increase in the predator's death rate $m_e$ and decrease in the prey growth rate $r_e$, respectively.

\bigskip
\begin{figure*}[tbh]
 \centering
\includegraphics[width=0.7\textwidth,angle=0]{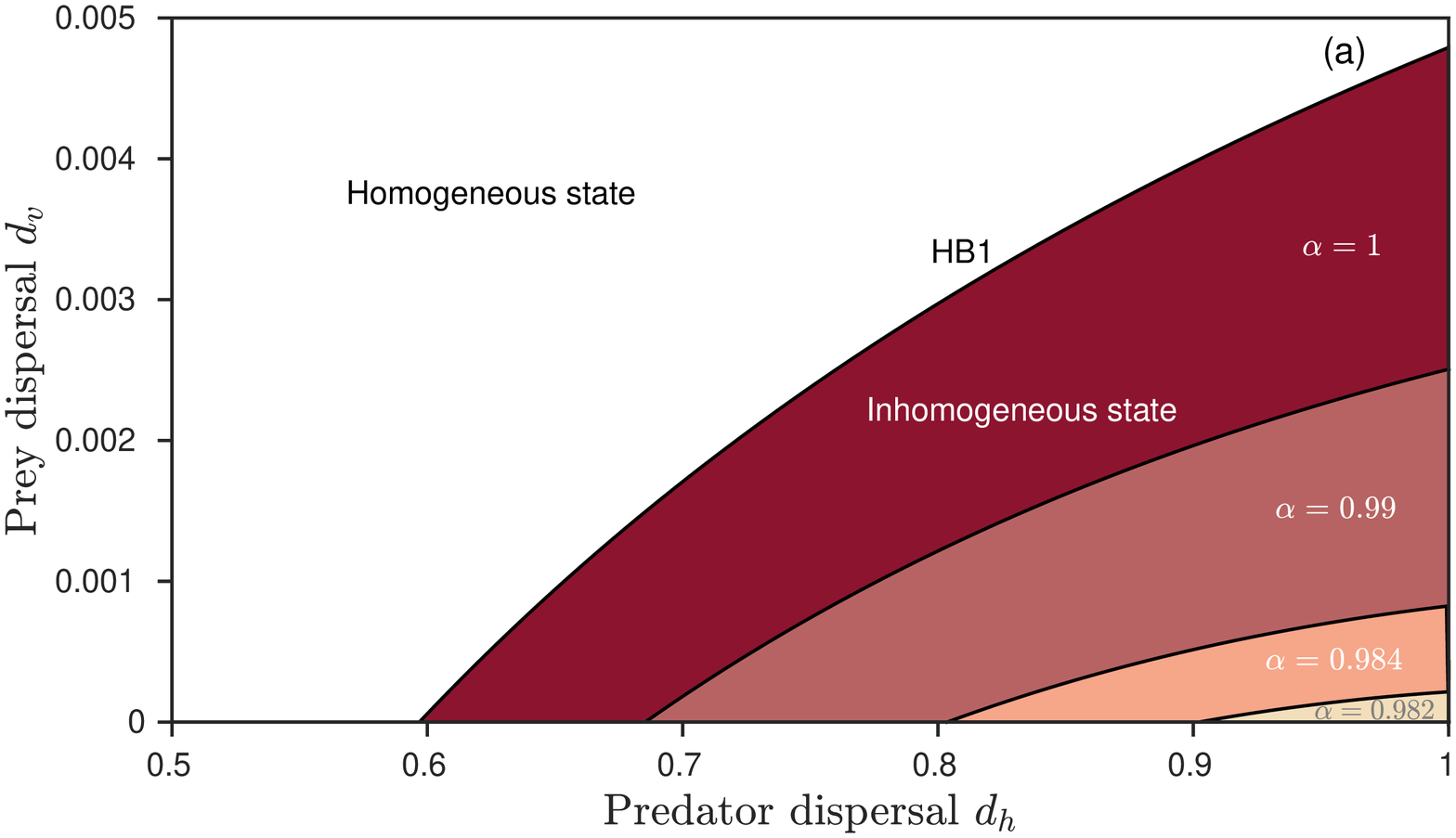} \\
\includegraphics[width=0.7\textwidth,angle=0]{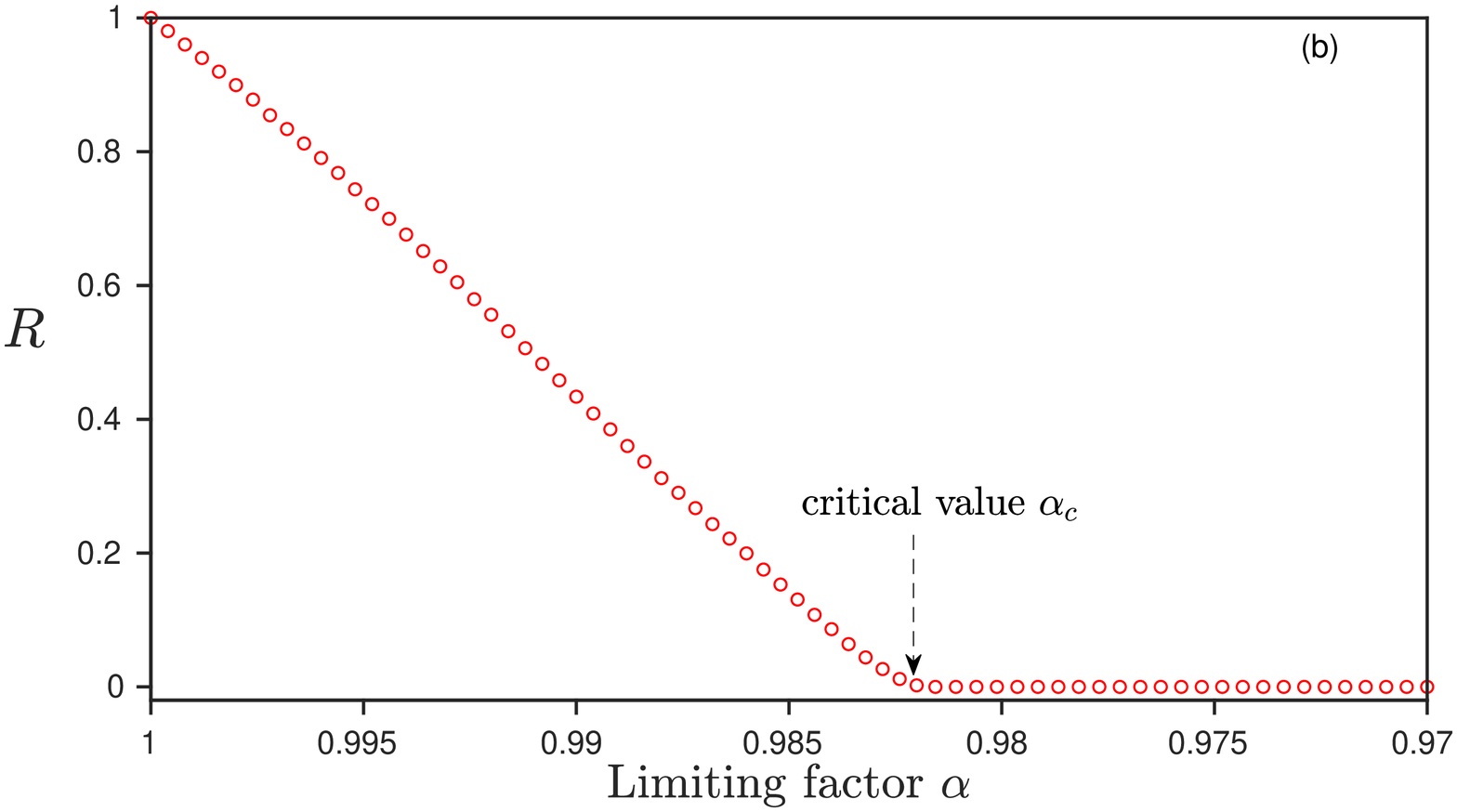} 
\caption{(a) Two parameter bifurcation diagram in $d_h-d_v$ space at  different $\alpha$ values: Hopf bifurcation (HB1) curve separates the region into homogeneous and inhomogeneous stable states. The region below the curve indicates the spread of the  inhomogeneous stable states that decreases for decreasing  $\alpha$. (b)  The normalized area of the spread of the stable inhomogeneous steady states defined as $R = S(\alpha)/S(\alpha = 1)$, where $S(\alpha)$ denotes the spread of the stable inhomogeneous steady states for $\alpha$. Other parameter values are $\beta=1$, $r=0.5$, $k=0.5$,   $a=1$, $c=0.5$, $b=0.16$ and $m=0.2$.}
  \label{fig3}
\end{figure*}
\section{Results}
\bigskip

Before 
unravelling the effect of limited dispersal, we illustrate various dynamical states admitted by the  metacommunity model  for distinct values of the system parameters in Fig.~\ref{fig1}.  
The limiting factors are fixed as $\alpha=\beta=1$, so that the coupling is the regular diffusive coupling, which accounts for the uncontrolled dispersal~\cite{ravkc2021}.
We numerically integrate the governing equation of the metacommunity model using the Runge-Kutta 4th order algorithm with a step size of $h=0.01$. 
We have fixed the  other parameters  as $r=0.5$, $a=1$, $c=0.5$, $b=0.16$,  $m=0.2$,  and $d_v=0.001$ throughout the manuscript unless otherwise specified. 
 Note that we have chosen the model parameters  in order to get oscillatory dynamics in the uncoupled model. Indeed, the uncoupled predator-prey model exhibits the oscillatory dynamics  when 
\[k  > \frac{b \left[a c+ m + (1 - \alpha)  d_h\right]}{a c+ (\alpha - 1)  d_h-m}. \]   The model (\ref{eq1})  exhibits  similar qualitative dynamics even for other choice of the parameters.
Homogeneous
oscillatory state is observed for the predator dispersal rate $d_h=0.2$ and for the carry capacity $k=0.2$ as depicted in Fig.~\ref{fig1}(a), which has  a high probability of extinction 
under environmental stochasticity during the epochs of low population density.   Phase synchrony (inhomogeneous oscillatory states) can be observed for  $d_h=1.0$ and
$k=1.2$ (see Fig.~\ref{fig1}(b)), which is also prone to global extinction like the homogeneous oscillatory state  in Fig.~\ref{fig1}(a), but with  a lesser probability when compared to
the latter at low population density.  Anti-phase synchronization  (inhomogeneous oscillatory states)   can be observed 
in  Fig.~\ref{fig1}(c) for $d_h=0.001$ and $k=0.6$, which has a much lesser chance of extinction when compared to the homogeneous and phase synchronized
 populations. One can also observe inhomogeneous steady states (see  Fig.~\ref{fig1}(d) for $d_h=0.45$ and $k=0.6$) of the populations, which usually coexist with
homogeneous (synchronized) oscillatory state, resulting in alternative dynamical states  for the population to promote their persistence.

 \subsection{limiting predator dispersal along with spatial interaction}
In this section, we will unravel the effect of limiting the predator dispersal  on the dynamical transitions of  the metacommunity model  as a function of
spatial parameters (predator and prey dispersal rates).  
We fix $\beta=1$, so that the preys are allowed to disperse completely, whereas only the  predator dispersal is controlled by decreasing the
value of $\alpha$. 
Bifurcation diagrams, plotted using XPPAUT~ \cite{Xpp02}, illustrating the dynamical transitions
as a function of the predator dispersal rate are depicted in  Fig.~\ref{fig2} for the carrying capacity $k=0.5$.
Extrema of  stable homogeneous oscillatory state are plotted as filled circles, while that of unstable oscillatory 
states are represented by unfilled circles. Stable and unstable steady states are  indicated by solid and dashed curves/lines, respectively.  
For $\alpha=1$, the  dispersal among the metacommunity is
 the usual diffusive coupling between the patches.  Only stable homogeneous (synchronized) oscillation among the  populations is observed all along 
the low values of the predatory dispersal  for  $\alpha=1$, where the homogeneous 
steady state is unstable (see Fig.~\ref{fig2}(a)).  The former prevails in the entire  explored range of the predator dispersal rate.
Subcritical pitchfork bifurcation onsets at $d_h=0.48$  resulting in the coexistence of unstable homogeneous
and inhomogeneous steady states along with the homogeneous oscillatory state in the range of $d_h\in[0.48, 0.597)$.  A Hopf bifurcation at $d_h=0.597$ leads to
the onset of stable inhomogeneous steady (asynchronous) states, which coexist along with  unstable inhomogeneous oscillatory states.  The onset of the former  gives rise to
the emergence of alternative stable states of the populations thereby leading to an increase in the degree of persistence of the metacommunity in the range of $d_h\in[0.597, 1.0]$.

Now, we control the dispersal rate of the predators using the limiting factor $\alpha$ and  investigate the effect of limited dispersal on the metacommunity persistence.  The dynamical
transitions of the prey density are depicted in Fig.~\ref{fig2}(b) for $\alpha=0.985$.  It is evident from the figure that even a feeble decrease in the limiting factor, that is a negligible 
fraction of decrease in dispersal,  results in drastic changes in the metacommunity dynamics. The Hopf bifurcation, resulted in the onset of alternative stable states, 
now emerges at  a higher dispersal rate at $d_h=0.7745$ reducing the tolerance of the metacommunity persistence to a narrow range of predator dispersal rate. 
This effect of limiting dispersal increases to a larger degree for further negligible decrease in  the fraction of dispersal.  As a consequence,  the  alternative stable states
lose its stability in the explored range of $d_h$ even for $\alpha=0.98$, denoting a very small decrease in the degree of the dispersal,  endangering the persistence of the
 metacommunity  due to the presence of  the homogeneous (synchronous)
oscillatory state as the only dynamical state of the metacommunity, which is highly prone to extinction. In the following, we will explore the effect of controlling
the predator dispersal on the metacommunity dynamics as a function of the dispersal rate of both  prey and predator populations.
 
  \begin{figure*}[tbh]
 \centering
\includegraphics[width=0.7\textwidth,angle=0]{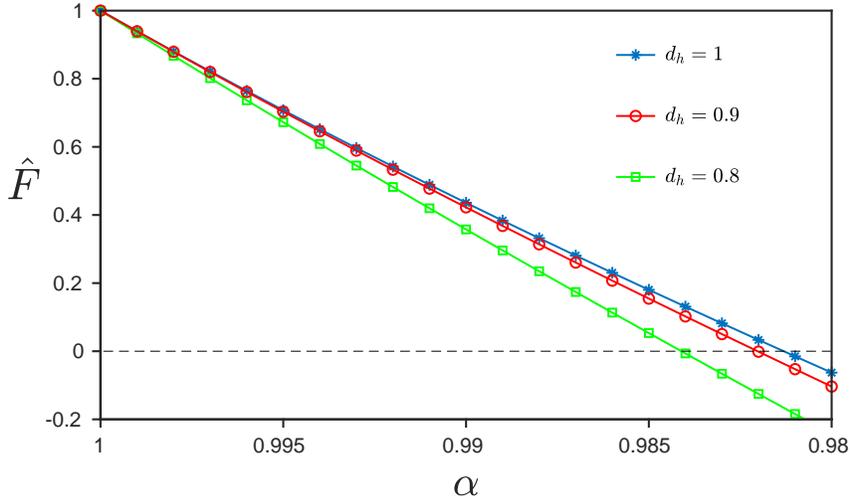}\\
\caption{Analytical stability curve for different predator dispersal rates.  The values of the parameters are the same as in Fig.~\ref{fig3}.}
  \label{fig3a}
\end{figure*} 
 
We have depicted the spread of the inhomogeneous steady (asynchronous, alternative) states in the two-parameter bifurcation diagram in Fig.~\ref{fig3}(a) for different degree of
the dispersal by reducing the limiting factor $\alpha$.  Homogeneous oscillatory state is  stable in the entire parameter space, while the  inhomogeneous steady states  are stable
in the shaded region, where both the former and the latter coexists.
The transition from homogeneous to inhomogeneous steady state is via the Hopf bifurcation as depicted in  one parameter bifurcation diagrams in Fig.~\ref{fig2}. 
Note that the Hopf bifurcation curves are obtained from XPPAUT. 
The spread of the inhomogeneous steady states is maximum for $\alpha=1$, which monotonically decreases for further decrease in the degree of the predator dispersal.  Eventually,
the inhomogeneous steady states  destabilized  completely in the same parameter space, where it was stable  earlier,  at a critical value of $\alpha$ resulting  in  the
homogeneous (synchronous) oscillatory state as the only dynamical state of the metacommunity.  The normalized area of the spread of the stable inhomogeneous steady states
defined as $R=S(\alpha)/S(\alpha=1)$, where $S(\alpha)$ denotes the spread of  the stable inhomogeneous steady states for $\alpha$, is depicted in Fig.~\ref{fig3}(b)  as
a function of $\alpha$.  It is evident from the figure that  the spread of the  stable inhomogeneous steady states decreases monotonically and eventually vanishes
at the critical value of  $\alpha=\alpha_{c}\approx 0.9818$ corroborating the dynamical transitions observed in the one-  and two-parameter bifurcation diagrams.

\begin{figure*}[tbh]
 \centering
\includegraphics[width=0.7\textwidth,angle=0]{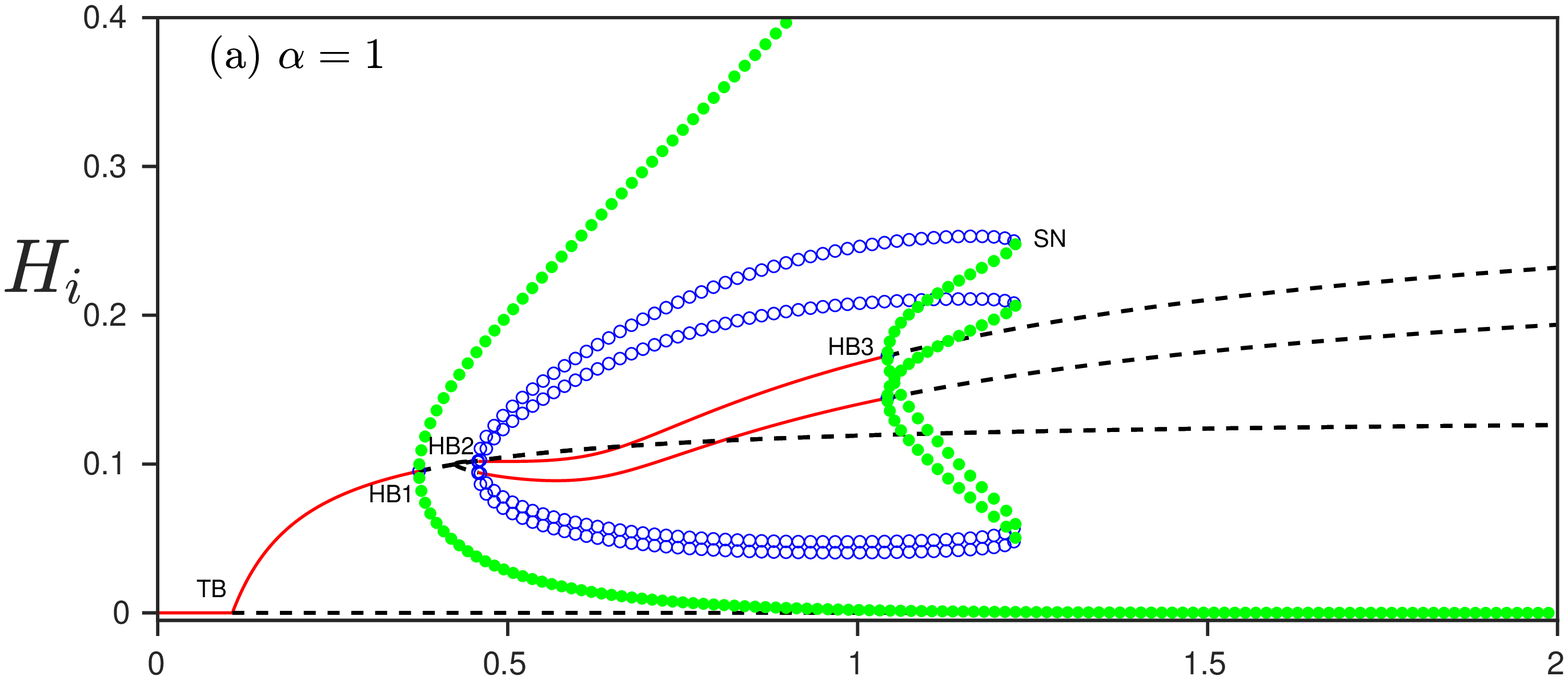}\\
\includegraphics[width=0.7\textwidth,angle=0]{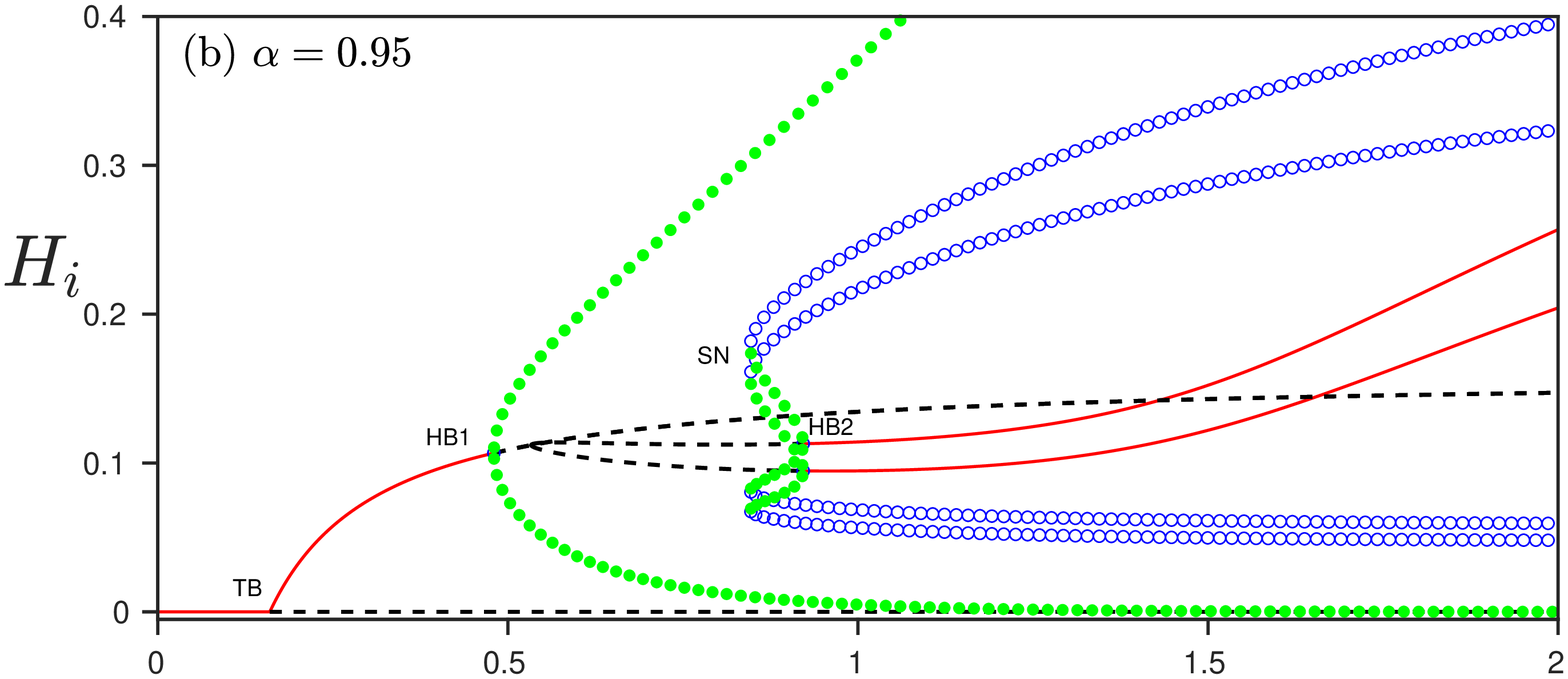}\\
\includegraphics[width=0.7\textwidth,angle=0]{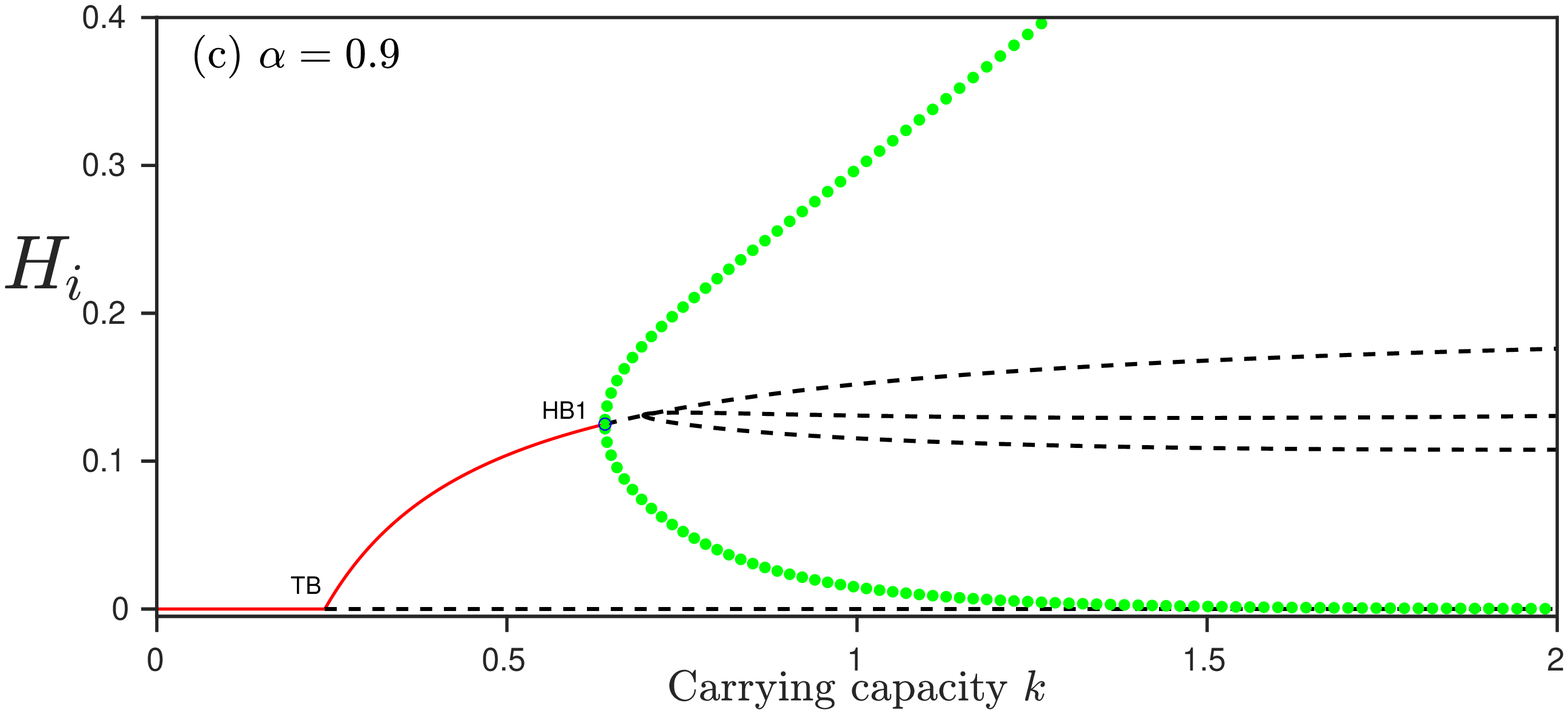}
\caption{Metacommunity dynamics for limiting the predator dispersal: Bifurcation diagram as a function of the carrying  capacity ($k$) is shown for different $\alpha$ values.  Homogeneous and inhomogeneous states of the metacommunity are shown for  (a) $\alpha=1$, (b) $\alpha=0.95$, and (c) $\alpha=0.9$. For a  decrease in $\alpha$,  inhomogeneous stable steady states (red solid lines) becomes unstable.  Here  TB, HB and SN correspond to  transcritical bifurcation, Hopf bifurcation
  and saddle-node bifurcation, respectively. Other parameters are $r=0.5$,
  $a=1$, $c=0.5$, $b=0.16$, $m=0.2$, $d_v=0.001$ and $d_h=1$.}
  \label{fig4}
\end{figure*}

It is  extremely difficult to extract the inhomogeneous steady states and deduce its stability condition analytically for finite values of prey dispersal rate $d_v$. 
However, for $d_v=0$, the inhomogeneous steady states $(V_1^\ast, H_1^\ast, V_2^\ast, H_2^\ast)$ can be deduced as
\begin{widetext}
\begin{align*}
V_1^\ast&=\frac{q_1+\sqrt{q_3-q_1^2}}{q_2}\;,\\
V_2^\ast&=\frac{-\sqrt{q_4^2-4 q_5 \left(\alpha  b k d_h-b k d_h-\alpha  b V_1^\ast d_h-b k m+\alpha  k V_1^\ast d_h-\alpha  V_1^{^\ast 2} d_h\right)}+q_4}{2 q_5}\;,\\
H_1^\ast&=-\frac{\left(b+V_1^\ast\right) \left(k V_1^\ast d_v-k V_2^\ast d_v-k r V_1^\ast+r V_1^{\ast 2}\right)}{a k V_1^\ast}\;,\\
H_2^\ast&=-\frac{\left(b+V_2^\ast\right) \left(k V_2^\ast d_v-k V_1^\ast d_v-k r V_2^\ast+r V_2^{\ast 2}\right)}{a k V_2^\ast}\;,
\end{align*}
where, 
\begin{align*}
q_1&=\left(d_h (2 m-2 a c)+(m-a c)^2-\left(\alpha ^2-1\right) d_h^2\right) \left(a c k+(\alpha +1) (b-k) d_h+m (b-k)\right), \\
 q_2&=2 \left(a c+(\alpha -1) d_h-m\right) \left(a c-(\alpha +1) d_h-m\right){}^2,\\
 q_3&=q_2 \left(a c-(\alpha +1) d_h-m\right)\left(a^2 c^2 k^2+2 d_h \left(a c k (\alpha  b-k) +m \left(b^2+k^2\right)\right)-2 a c k^2 m-\left(\alpha ^2-1\right) \left(b^2+k^2\right) d_h^2+m^2 \left(b^2+k^2\right)\right),\\
q_4&=-a c k-b d_h-b m+k d_h+k m, \qquad  \text{and} \\
q_5&=\left(-a c+d_h+m\right).
\end{align*}

The corresponding  Jacobian  matrix can be given as
\begin{center}
	\(A=\left( \begin{array}{cccc}
 -d_v+r \left(1-\frac{2 V_1^\ast}{k}\right) +s_1-s_2 & -s_5 & d_v & 0 \\
 c s_2-c s_1 & c s_5-d_h -m & 0 & \alpha  d_h \\
 d_v & 0 & -d_v+r \left(1-\frac{2 V_2^\ast}{k}\right) +s_3-s_4 & -s_6 \\
 0 & \alpha  d_h & c s_4-c s_3 & c s_6-d_h-m  \end{array} \right),\) 
\end{center}
\end{widetext}
where, 
$s_1=\frac{a h_1 V_1^\ast}{(b+V_1^\ast)^2}$,  $s_2=\frac{a h_1}{b+V_1^\ast}$, $s_3=\frac{a h_2 V_2^\ast}{\left(b+V_2^\ast\right){}^2}$, $s_4=\frac{a h_2}{b+V_2^\ast}$, $s_5=\frac{a V_1^\ast}{b+V_1^\ast}$,  and $s_6=\frac{a V_2^\ast}{b+V_2^\ast}$.  One can deduce the characteristic equation from the Jacobian as 
\begin{align}
\lambda^4-Tr(A)\lambda^3+a_2\lambda^2-a_3\lambda+Det(A)=0.
\end{align}
The condition for the stability of the  inhomogeneous steady  (alternative) states can be obtained  in terms of the system parameters as
\begin{align}
F_{\alpha}(a,b,c,k,m,r,\alpha,d_h)&\,=a_3(Tr(A) a_2-a_3) \nonumber \\ 
&\,-Tr(A)^2Det(A)>0.
\end{align}
The inhomogeneous steady state is stable for $F_\alpha>0$ and unstable for $F_\alpha<0$.   $\hat{F}=F_\alpha/F_1$  is depicted  in Fig.~\ref{fig3a} as a function of the limiting factor 
for three predator dispersal rates $d_h=0.8, 0.9$ and $1$. The value of the limiting factor $\alpha$ at which $\hat{F}=0$ corresponds to the critical value 
 $\alpha=\alpha_c$, where there is a switch in the stability of the  inhomogeneous steady state.  The  Hopf bifurcation curves plotted using XPPAUT  in Fig.~\ref{fig3}(a) for the
 critical values of the limiting factor $\alpha_c=0.984$ and $0.982$ coincide with the predator dispersal rate ($x$-axis where $d_v=0$) nearly at  $d_h=0.8$ and $0.9$,
 respectively,  corroborating the critical value of the limiting factor, at which there is change in the stability of the  inhomogeneous steady state, obtained using the analytical stability curve in Fig.~\ref{fig3a}.

\begin{figure*}[tbh]
 \centering
\includegraphics[width=0.7\textwidth,angle=0]{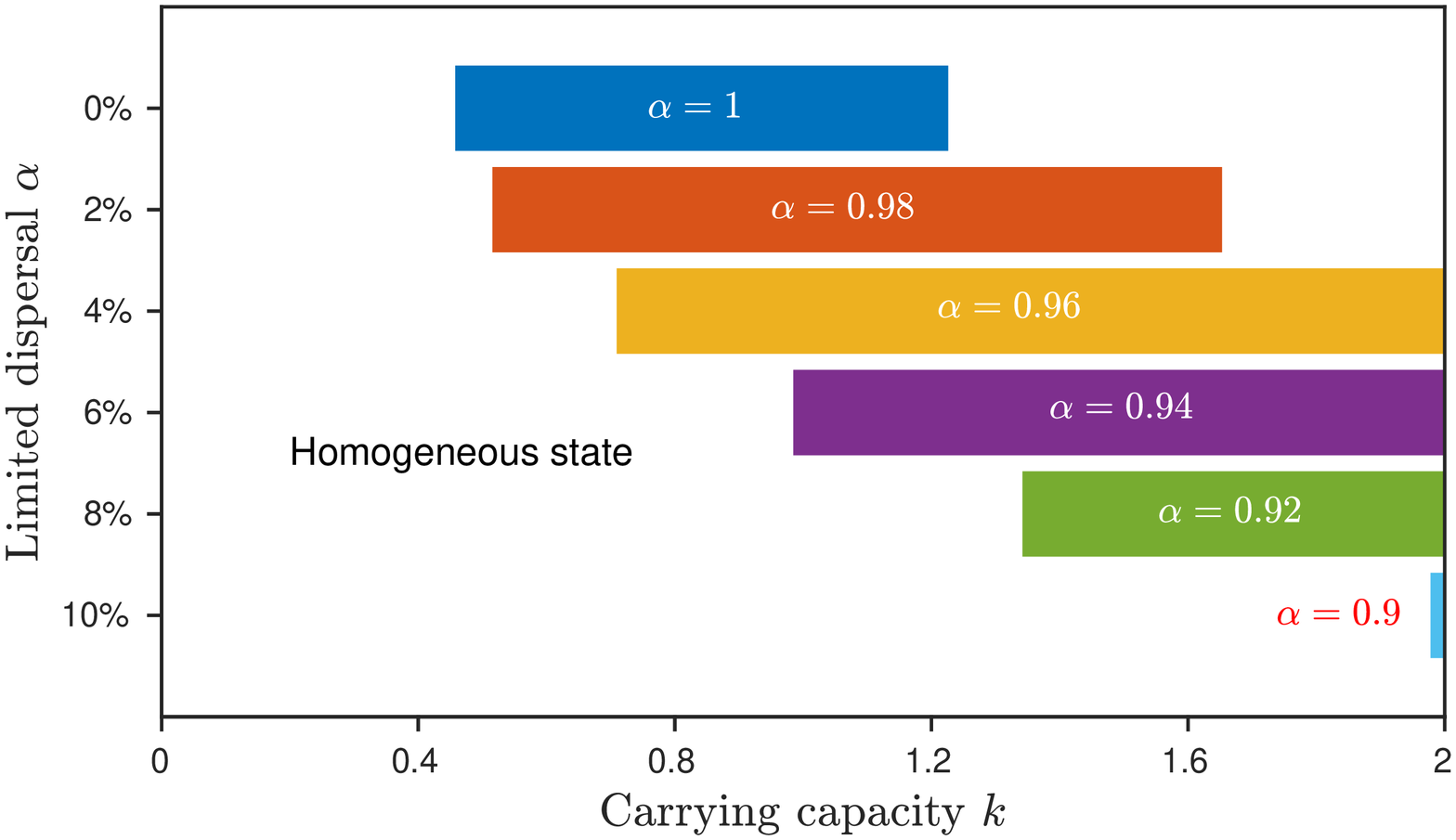} 
\caption{Inhomogeneous states of the metacommunity for different $\alpha$ values: Each horizontal bar denotes the parameter range of $k$ where the metacommunity exhibits  inhomogeneous stable  steady (asynchronous alternative) states. Other parameters are $\beta=1, r=0.5$, $a=1$, $c=0.5$, $b=0.16$, $m=0.2$, $d_v=0.001$ and $d_h=1$.}
  \label{fig5}
\end{figure*}
\subsection{limiting predator  dispersal along with local interaction}

\bigskip

In this section,  we investigate the influence of the limited  predator dispersal and  a local parameter (carrying capacity) on the dynamics of the metacommunity 
for  $\beta=1$. The predator density is 
depicted in the bifurcation diagram as a function of the carrying capacity $k$ (see Fig.~\ref{fig4}).  The prey and predator dispersal rates are fixed as $d_v=0.001$ and
$d_h=1$, respectively.  The  dynamical states are the same as in Fig.~\ref{fig2}.   The trivial steady state 
lose its stability via  a transcritical bifurcation  at $k= k_{TB}$  leading to a nontrivial homogeneous steady state, which
undergoes  a supercritical Hopf bifurcation (HB1)  resulting in stable homogeneous oscillation and  an unstable steady state at $k=0.4$ for 
$\alpha=1$ (see Fig.~\ref{fig4}(a)). Using the standard linear stability analysis around the homogeneous steady state $\left(\frac{kr+kd_v(\beta-1)}{r},0,\frac{kr+kd_v(\beta-1)}{r},0\right)$,
the  transcritical bifurcation  at $k= k_{TB}$  can be deduced as  \[  k_{TB} = \frac{b (d_h +m  - \alpha   d_h) }{ a c+ d_h(\alpha-1)-m }.\]  Similarly, 
one can obtain  the Hopf bifurcation  \[k_{HB} = \frac{b \left[a c+ m + (1 - \alpha)  d_h\right]}{a c+ (\alpha - 1)  d_h-m},\]  by performing   a linear stability analysis around the 
 homogeneous steady state $(V^*, H^*, V^*, H^*)$, where $V^*= \frac{bm+bd_h(1-\alpha)}{ac-m-d_h(1-\alpha)}$, $H^* = \frac{bc\left[ack(r+d_v(\beta-1))-\left(br+kr+kd_v(\beta-1)\right)\left(m+d_h(1-\alpha)\right) \right]}{k(ac-m+d_h(\alpha-1))^2}$.  Note that for $\beta=\alpha=1$, the Hopf and pitchfork bifurcation points/curves can be reduced to
 that in Ref.~\cite{ravkc2021}.

\bigskip
\begin{figure*}[tbh]
 \centering
\includegraphics[width=0.7\textwidth,angle=0]{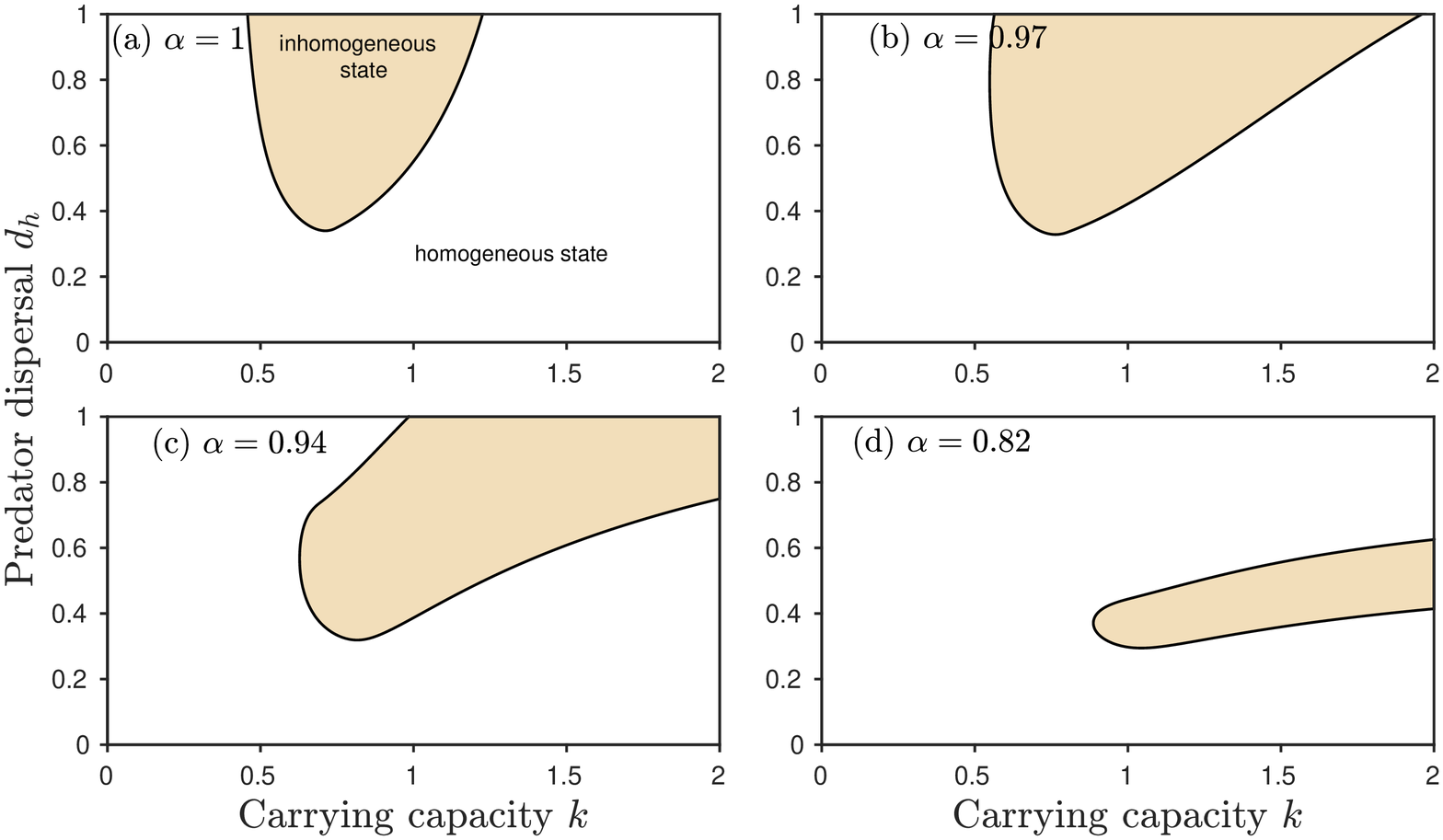} 
\caption{Two parameter bifurcation diagram in $(k, d_h)$  parameter space for different $\alpha$ values: Here the shaded region  indicates the parameter space where the metacommunity exhibits inhomogeneous stable states. Outside the shaded region, metacommunity exhibits only homogeneous state. For decreasing $\alpha$, inhomogeneous region is increased at first, and then decreased.  Other parameters are $r=0.5$,   $\alpha=1$, $\beta=0.5$, $B=0.16$, $m=0.2$ and $d_v=0.001$.}
  \label{fig6}
\end{figure*}

The stable 
homogeneous oscillatory state prevails in the entire explored range of $k\in [0.4, 2]$.  Subcritical Hopf bifurcation (HB2)  emerges at $k=0.458$  resulting in
unstable inhomogeneous oscillations and stable inhomogeneous steady states in the range of $k\in[0.458, 1.043)$ leading to the existence of alternative stable states
of the metacommunity.   The latter lose its stability via a supercritical Hopf bifurcation (HB3) at $k=1.043$ resulting in stable  inhomogeneous oscillations in the range  $k\in[1.043, 1.226)$  and 
unstable inhomogeneous steady states in the range  $k\in[1.043, 2.0]$, leaving behind the synchronized oscillatory states as the only stable state of the metacommunity  
A saddle-node bifurcation emerges at $k=1.226$. In the following, we will unravel the effect of limited dispersal as a function of the carrying capacity.

The bifurcation diagram elucidating the dynamical transitions of the metacommunity for $\alpha=0.95$ is depicted in Fig.~\ref{fig4}(b).  The dynamical states and their
transitions are almost similar to that discussed in Fig.~\ref{fig4}(a) for $\alpha=1.0$.  Nevertheless, it is evident from the figure that the spread of the stable inhomogeneous steady 
 (asynchronous alternative states) states  increased to a larger extent, thereby increasing the degree of persistence of the metacommunity. However, this tendency, enhancing
 the spread of stable inhomogeneous steady states, of the degree of the dispersal persists up to a critical value and then the steady states lose its stability eventually resulting
 in the homogeneous (synchronous) oscillatory state as the only dynamical state as illustrated in  Fig.~\ref{fig4}(c) for $\alpha=0.9$.  The spread of the stable
 inhomogeneous steady states  for different values of $\alpha$  is depicted in   Fig.~\ref{fig5} as a function of the carrying capacity $k$. It is also evident from the figure that
 the degree of the spread  of  inhomogeneous steady states   increases initially for a small decrease in the dispersal and eventually  decreases  for further decrease in 
 the dispersal beyond a critical value of $\alpha$.  
 
  \bigskip
 \begin{figure*}[tbh]
 \centering
\includegraphics[width=0.7\textwidth,angle=0]{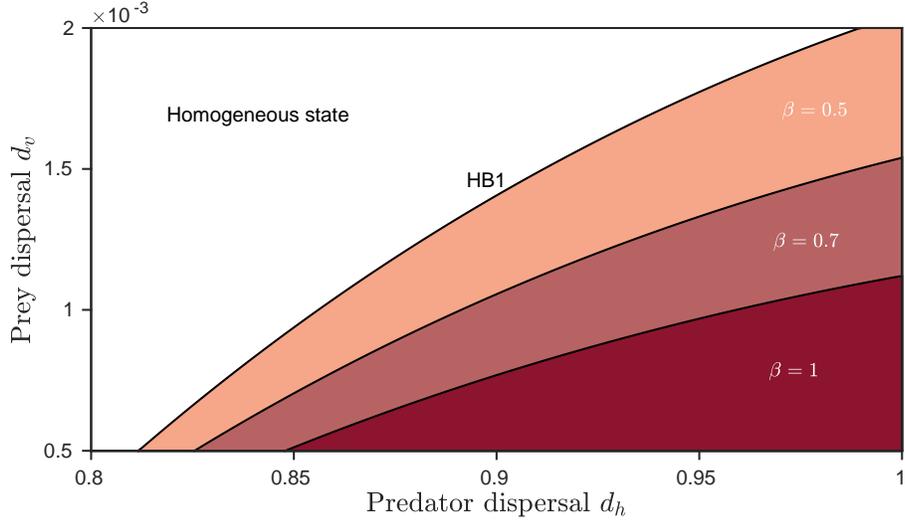} \\
\caption{Two parameter bifurcation diagram in $d_h-d_v$ space at  different $\beta$ values: Hopf bifurcation (HB1) curve separates the region into homogeneous and inhomogeneous stable states. Other parameter values are $r=0.5$, $k=0.5$,   $a=1$, $c=0.5$, $b=0.16$,  $m=0.2$ and $\alpha=0.985$.}
  \label{fig6a}
\end{figure*}

 The spread of the homogeneous and inhomogeneous states are shown in Fig.~\ref{fig6}  as two parameter bifurcation diagrams in  $(k, d_h)$ parameter space
 for different degree of the predator dispersal.  The homogeneous state prevails in the entire parameter space, where as the stable inhomogeneous steady states prevail only in the
 shaded regions leading to multistability.  It is also evident from the figures that the spread of the stable inhomogeneous steady states increases for a small decrease
  in the degree of  the predator dispersal 
 up to a critical value of $\alpha$ and then it starts decreasing as a function of $\alpha$.   The stable inhomogeneous steady states eventually destabilized from 
 the entire parameter space, where it was stable,
 resulting in the homogeneous states as the only dynamical states of the metacommunity.  Thus, the degree of limiting the predator dispersal  results in increasing the
 persistence of the metacommuntiy  until a critical value, above which it is endangered by the presence of homogeneous (synchronous) states as the only
 dynamical states of the metacommunity.

 \bigskip
\subsection{limiting prey  dispersal}
 \bigskip
 Now, we will investigate the effect of limiting the prey dispersal on the metacommunity dynamics as a function of the spatial parameter.  
 For complete dispersal of the predators, that is for $\alpha=1$,   
 the effect of  limiting the prey dispersal  on the metacommunity dynamics can be seen only in a narrow range of  the parameter space. Hence,
 we have fixed $\alpha=0.985$, and depicted the two-parameter bifurcation diagram  in the $(d_h, d_v)$  parameter space, where the effect of limiting
 the prey dispersal is well pronounced as evident from Fig.~\ref{fig6a}.  Homogeneous oscillatory state is stable in the entire parameter space, while the
 inhomogeneous steady states are stable only  in the shaded region. The transition from homogeneous oscillator state to inhomogeneous steady states
 is via the transcritical Hopf bifurcation (HB1) curve, which is obtained from XPPAUT.  It is to be noted that  for $\beta=1$, that is for uncontrolled prey dispersal,
 the spread of the  inhomogeneous steady states is rather limited to  smaller  values of  $d_v$.  However,  decreasing $\beta$, that is  limiting
 the prey dispersal,  manifests the stable inhomogeneous steady states in a larger region of the $(d_h, d_v)$  parameter space.
 For instance, the spread of the stable inhomogeneous steady states are depicted  in Fig.~\ref{fig6a} for $\beta=0.7$ and $0.5$. 
 Thus,  limiting the prey dispersal  results in a large degree of the metacommunity persistence in contrast to the effect of limiting the predator dispersal. 
 It is also to be noted that for different degrees of local parameter (carrying capacity),  limiting the prey dispersal  results in an
 increase in the metacommunity persistence.
 
 \begin{figure*}[tbh]
 \centering
\includegraphics[width=0.85\textwidth,angle=0]{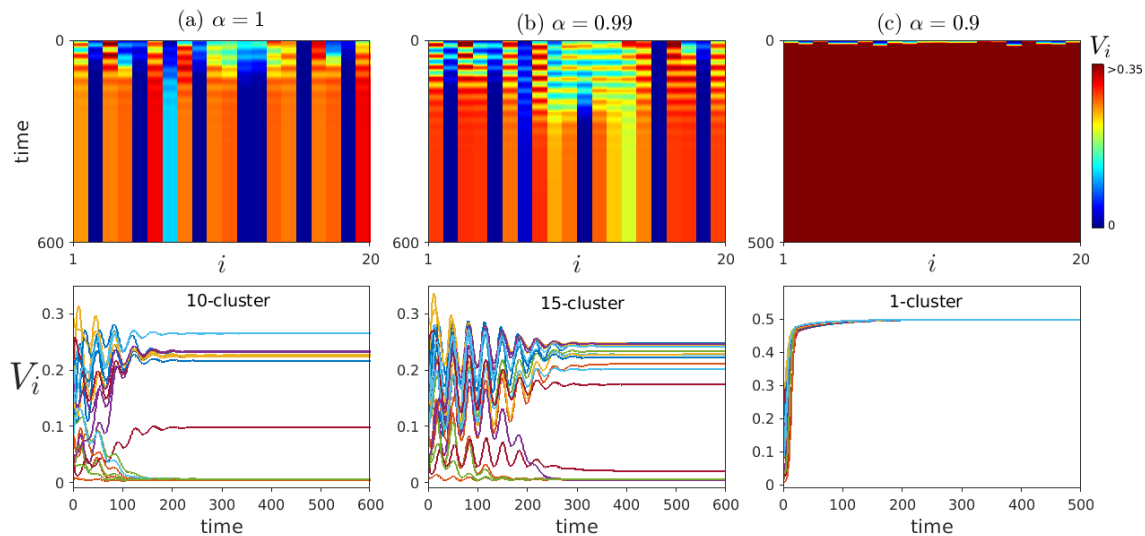} 
\caption{Spatio-temporal dynamics  of a random (Erdos-Renyi) network consisting of 20 patches:  Top panel shows the spatial dynamics of  the metacommunity, whereas the bottom panel shows the respective time series. The metacommunity shows dynamical transitions from 10 clusters to  15 clusters, which further organized into a single cluster
for decreasing  degree of the limiting factor ($\alpha$). Here initial conditions are the same for each $\alpha$ value. Other parameters are  fixed as $\beta=1$, $r=0.5$,  $a=1$, $c=0.5$, $b=0.16$, $m=0.2$ and $d_v=0.001$.}
  \label{fig7}
\end{figure*}
 \bigskip
\subsection{limiting predator dispersal in a larger network}
\bigskip

 To elucidate the generic nature of our results, we consider an arbitrary network consisting of $N=20$ patches, whose  governing equation is represented as
 \begin{subequations}\label{eq2}
  \begin{align}
  \frac{dV_{i}}{dt} &= rV_{i} \left(1-\frac{V_{i}}{k} \right)-
  \frac{a  V_{i}}{V_{i}+b} H_{i} +\frac{d_v}{d_j}\sum_{j=1}^Ng_{ij}\left(\beta V_j - V_i\right)\;,\\ \frac{dH_{i}}{dt} &=
   \frac{c a V_{i}}{V_{i}+b} H_{i} - m H_{i} + \frac{d_h}{d_j}\sum_{j=1}^Ng_{ij}\left(\alpha H_j -
  H_{i}\right)\;, 
  \end{align}
  \end{subequations}
 where $i=1,\ldots, N$.  $g_{ij}$ encodes the topology of the underlying network. $g_{ij}=g_{ji}=1$, if $i$th and $j$th oscillators are connected, otherwise $g_{ij}=g_{ji}=0$.
 $d_j=\sum_{j=1}^{N}g_{ij}$ corresponds to the degree of the $j$th oscillator.
 The other parameters are the same as in Eq.~(\ref{eq1}).  We have fixed the values of the parameters as
$\beta=1.0, r=0.5$,  $a=1$, $c=0.5$, $b=0.16$, $m=0.2, d_h=1$ $k=0.5$ and $d_v=0.001$.
 The dynamics of a  random (Erdos-Renyi)  network is depicted  as  spatiotemporal and time series plots in 
 Fig.~\ref{fig7} for different degree of the predator dispersal.  The network exhibits  a ten  cluster state, corresponding to ten stable steady states,
 for $\alpha=1$ as evident from the  spatiotemporal  and time series plots in Fig.~\ref{fig7}(a). Nevertheless, decreasing the degree of  dispersal to $\alpha=0.99$
 results in increase in the number of clusters (see Fig.~\ref{fig7}(b)).    Note that the multi-cluster state corresponds to the alternative states of
 the metacommunity elucidating a high degree of their persistence.  However, further decrease in the  degree of the dispersal decreases the number of
 multi-cluster and eventually manifests as a single cluster state,  as depicted in  Fig.~\ref{fig7}(c) for $\alpha=0.9$, at a critical value of $\alpha$
  signaling a low degree of  the metacommunity persistence.
  
\begin{figure*}[tbh]
 \centering
\includegraphics[width=0.7\textwidth,angle=0]{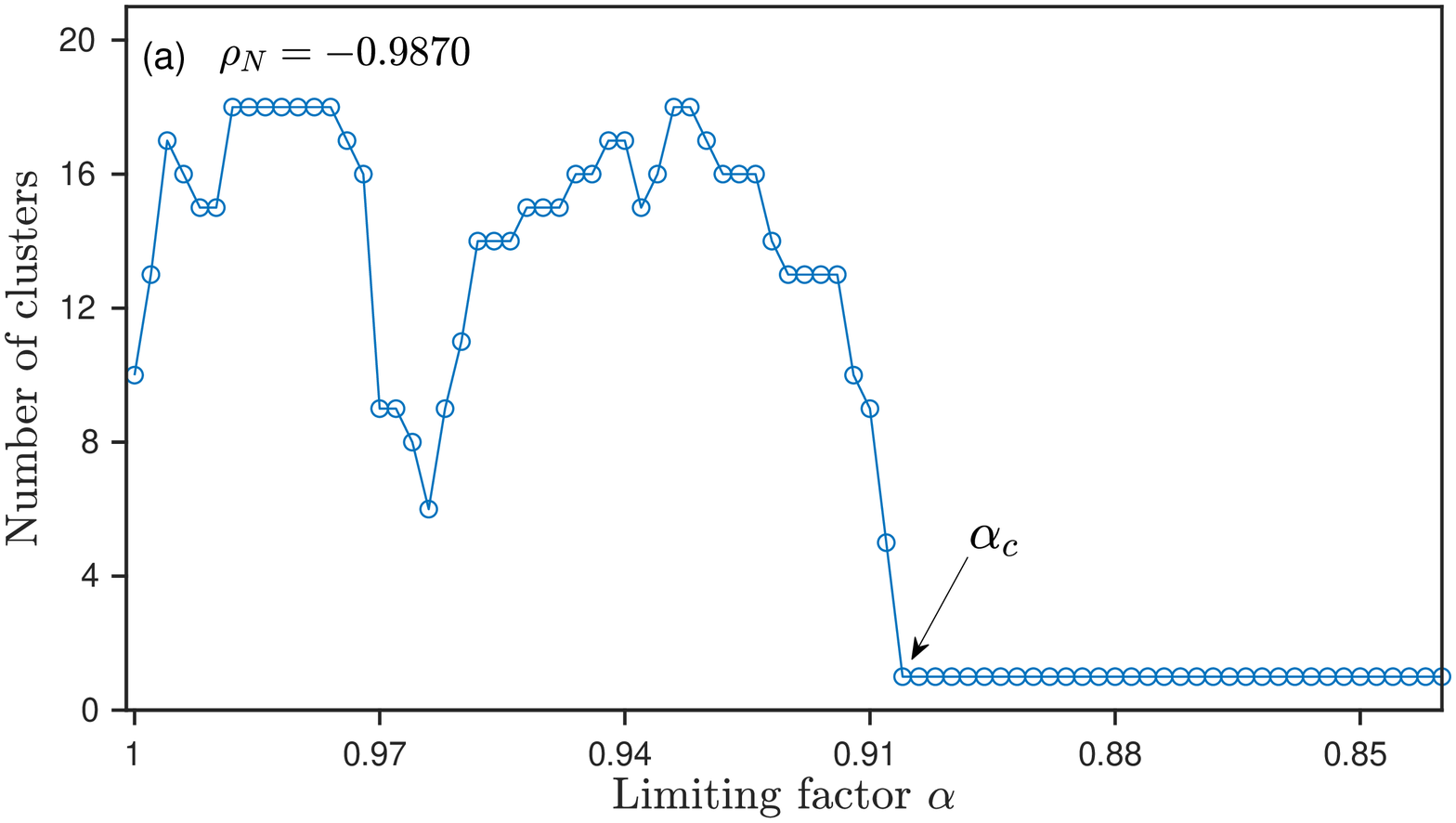}\\
\includegraphics[width=0.7\textwidth,angle=0]{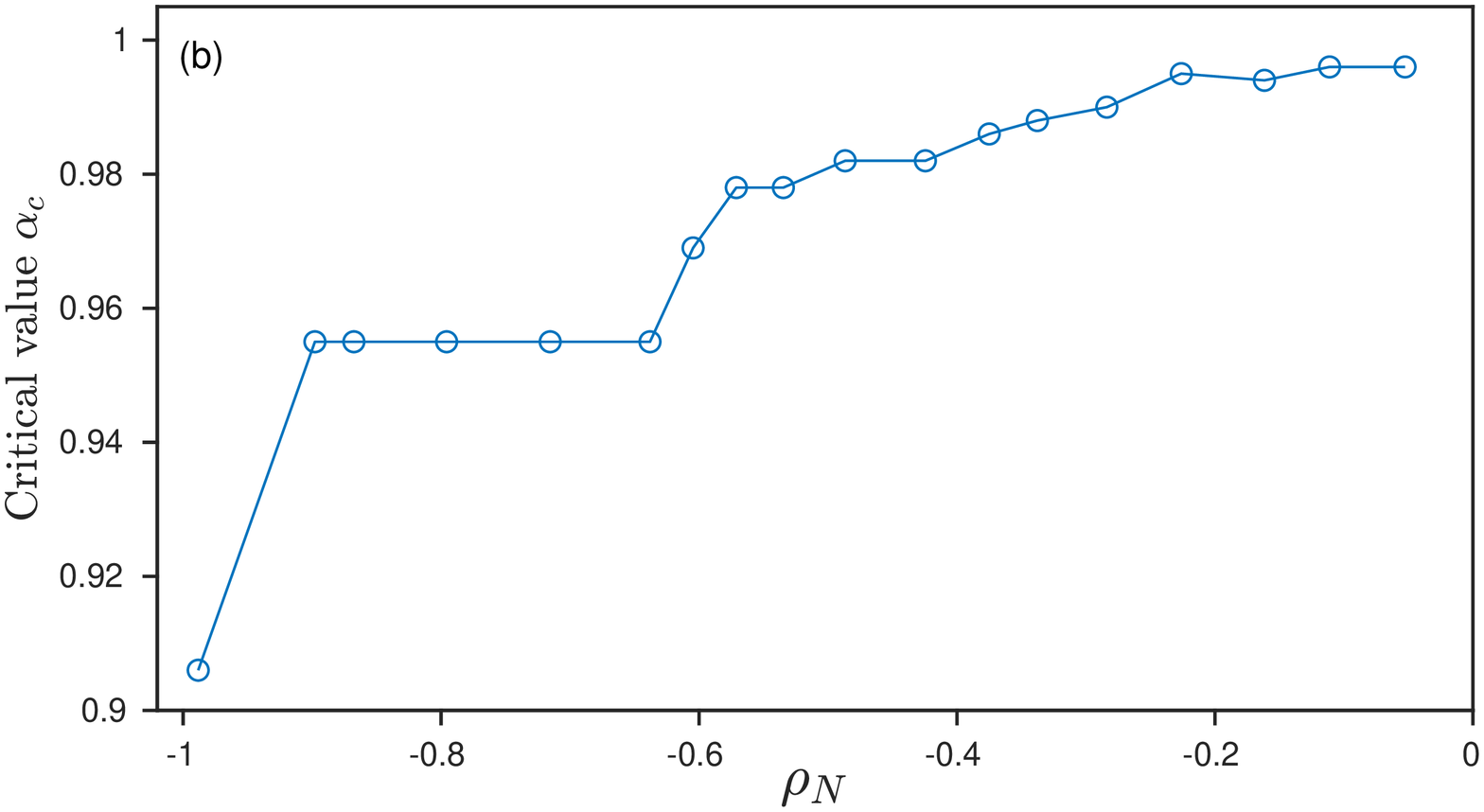}
\caption{(a) Number of clusters as a function of the limiting factor $\alpha$ for a random network of $20$ nodes showing the transition from inhomogeneous steady states
to homogeneous steady state, and (b) $\alpha_c$ as a function of the least bounding eigenvalue  $\rho_N\in[-1,0]$ of the connectivity matrix $g_{ij}$, characterizing 
the entire class of  complex networks}.
  \label{fig8}
\end{figure*}

The number of clusters is depicted as a function of the limiting factor $\alpha$ in  Fig.~\ref{fig8}(a) for the same values of the parameters as in  Fig.~\ref{fig7}.  
The size of the multi-cluster states initially varies for small decrease in the limiting factor and eventually decreases  resulting in  a single cluster state (homogeneous state)
at a critical value of $\alpha=\alpha_c=0.907$.  
Thus, it is evident that the observed results of the two patch metacommunity hold for a random network  of metacommunity thereby corroborating the generic nature of the results.

Denoting $\rho_j$'s ($j=1,2,\ldots, N$) as  the eigenvalues of  the connectivity matrix $g_{ij}$, characterizing an arbitrary network, which can be ordered as
$1.0=\rho_1\ge \rho_2\ge\ldots\ge-1/(N-1)\ge\rho_N\ge-1.0$~\cite{fma2006}.  The smallest bounding eigenvalue of the connectivity matrix  $g_{ij}$
 corresponding to the random network 
employed in Figs.~\ref{fig7} and~\ref{fig8}(a) is found to be $\rho_N=-0.987$.  It is known that the role of coupling topology on the stability of the homogeneous steady state
is determined by $\rho_N$~\cite{wzdvs2013,wzdvs2021}, which completely characterizes the effect of the connection topology corresponding to an arbitrary network.
We have depicted the  critical value of the limiting factor $\alpha_c$ as a function of  $\rho_N$~\cite{method}  in  Fig.~\ref{fig8}(b), which elucidates that our results are
generic to any arbitrary network.  Note that the entire range of  $\rho_N\in[-1,0]$  captures all possible coupling topologies. It is to be noted that limiting the prey dispersal rate always results in multi-cluster states (inhomogeneous steady state) promoting 
metacommunity persistence and hence the above analysis of calculating $\alpha_c$ cannot be applied to this case.

\begin{figure*}[tbh]
 \centering
\includegraphics[width=0.7\textwidth,angle=0]{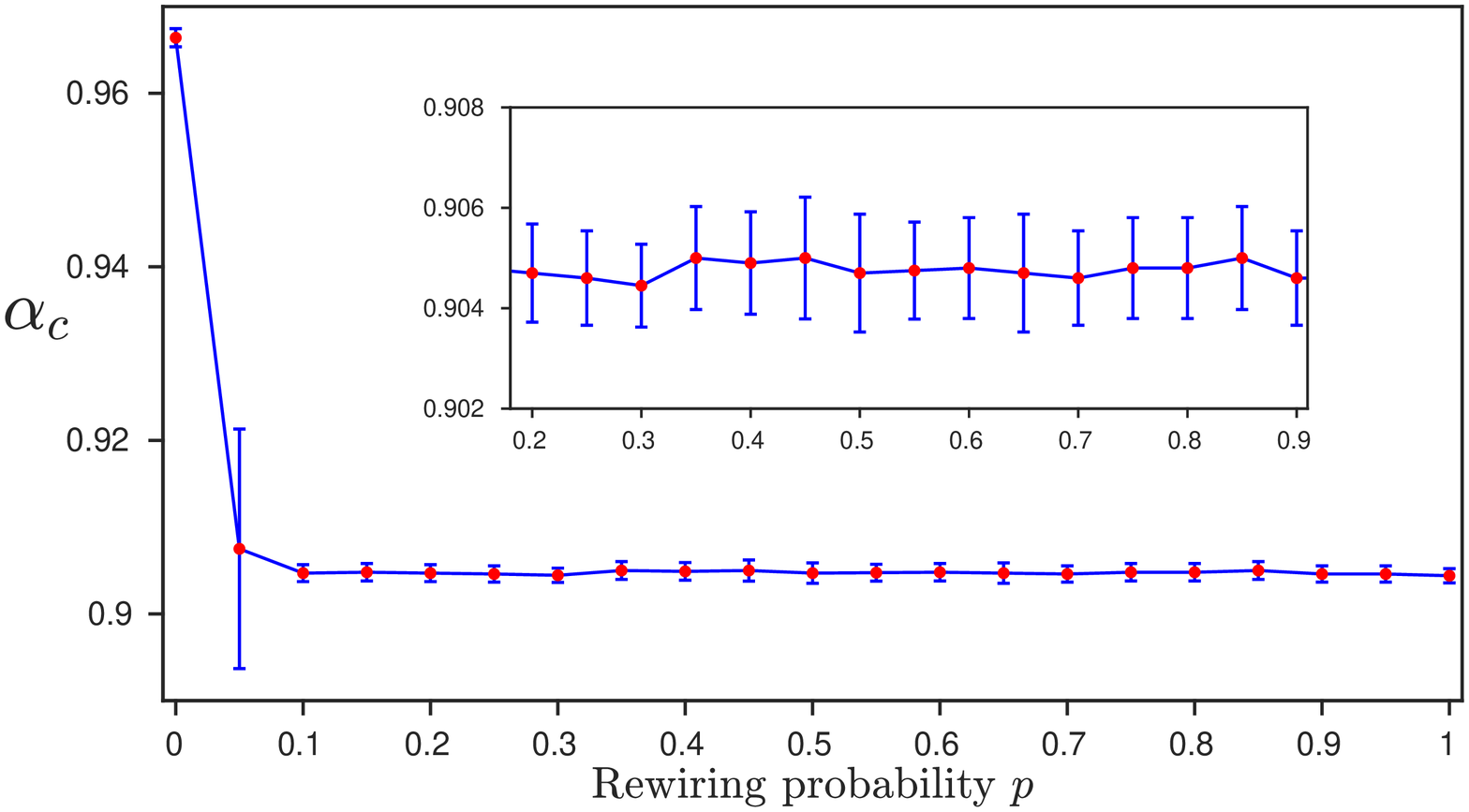}\\
\caption{ The critical value of predator  limiting factor $\alpha_c$  as a function of  the probability of rewiring $p\in[0,1]$
corresponding to the entire class of complex networks.
Mean of all the critical values of the limiting factor corresponding to each $N$  is indicated by the filled
circle for each $p$, while their variance is denoted by the error bars. 
 Inset  in the specific range of $p$ to clearly depict the error bars.}
  \label{fig11}
\end{figure*}

Further, different networks can be constructed for the same probability of rewiring `$p$'
from complex networks perspective. Noted that the range of $p\in[0,1]$ covers the entire class of
network topologies from regular, small world to random (scale-free) networks~\cite{wast98}.  
The critical value of $\alpha$  is depicted in Fig.~\ref{fig11} as a function of  $p$ with error bars characterizing the effect of various network sizes $N$.   
Specifically,  we have chosen $N=20$ to $500$ in steps of $10$. The mean of all the critical values of the limiting factor corresponding to each $N$  is indicated by the filled
circle for each $p$, while their variance is denoted by the error bars.   Inset in  Fig.~\ref{fig11}  clearly depicts the error bars. 
 The critical value of the limiting factor in the entire range of $p$  again corroborates the generic nature of our results.

\bigskip

\section{Discussion}

\bigskip

In this study, we have addressed how  limiting both predator and prey dispersal of a spatially distributed community affects its stability and persistence. Using  the homogeneous (synchronized) and inhomogeneous (asynchronized) dynamical states of  a diffusively coupled predator-prey metacommunity, we have elucidated the importance of controlled dispersal at  local and spatial scales. At spatial scale, a small decrease in the predator immigration through the limiting factor reduces the metacommunity persistence  by inducing the homogeneous state. On the other hand, at local scale, metacommunity persistence is increased for a small decrease in the predator immigration up to a critical value and then decreases for further increase in the degree of the limited predator dispersal. Moreover, our findings revealed that there exists a critical value for the limiting factor, below which  metacommunity persists and above which metacommunity  goes to  a high risk state.  However, limiting the prey dispersal promotes inhomogeneous steady states in a 
large region of the parameter space, thereby increasing the metacommunity persistence both at spatial and local scales.
Further, we  have showed the similar qualitative dynamics for  an entire class of complex networks 
 consisting of a large number of patches illustrating the robustness of our results, which
remains unaltered in a large range of model parameters.
 Thus, our findings revealed that dispersal-dependent responses strongly influence the metacommunity persistence. 
  Note that   the spread of homogeneous and inhomogeneous steady states in the one-parameter  bifurcation diagrams and 
 two phase diagrams only get rescaled for the rescaled model (\ref{eq1a}).

\bigskip
 
Dispersal is a fundamental process for structuring natural ecosystems  and it is widely recognized in
conservation and ecosystem management \cite{BlHu99,HoHa08}. Since dispersal can rescue local populations from complete extinction, it is  an important stabilizing mechanism for various ecological communities. However, dispersal induces synchrony among spatially connected populations which can elevate a high risk of extinction as compare to asynchronized dynamics \cite{BrHo04,LiKo04}. Hence, theoretical understanding of population synchrony controlled by dispersal received a great attention. Numerous theoretical studies have addressed the factors and mechanisms of synchrony using  various coupling strategies \cite{KeBj00,Ripa2000,GoHa08,RaDu16}.  Our study revealed the synchronized and asynchronized dynamics of the metacommunity  through a control on the diffusive dispersal. In particular, a decrease in the species immigration destabilizes the metacommunity through synchronized behaviour. Dispersal  stabilizes the metacommunity through a source-sink behaviour whereby some patches have a high population abundance and others have  a low population abundance \cite{Dia96}. Various inhomogeneous dynamical states shown in this study represent the source-sink behaviour, which in turn influences the metacommunity persistence.

\bigskip

``Paradox of enrichment", referring the existence of high-amplitude extinction-prone cycles due to increasing carrying capacity, can destabilize the ecological community \citep{Ros71}. Our  findings  on  varying  carrying capacity with a controlled dispersal elucidated the solution to the ``paradox of enrichment".  In particular, our  results are in line with previous studies that dispersal can prevent  the extinction prone cycles by creating inhomogeneous states and provide a solution to the paradox of enrichment \citep{Jan95,HaNa13,GoMo14}. In addition, our findings revealed that the  inhomogeneous states can be manifested as  homogeneous state by limiting the dispersal. Thus, limiting the dispersal offer insights to understand the stability and persistence of a spatially distributed community through synchronized and asynchronized dynamics.

 \bigskip
 
 In summary, this study presents the effect of limiting dispersal on  metacommunity persistence at local and spatial scales. The dynamics that we observe may apply to other  ecological scenarios such as environmental stochasticity, spatial heterogeneity, and higher tropical interactions (i.e., foodweb). Since these ecological scenarios increase the  complexity of system, a further investigation is required to understand how limiting the dispersal influences the stability of ecological communities. Overall, this study offers new insights on the stability and persistence of a spatially distributed ecological community.   \\

\bigskip
\section{Acknowledgements}
S. R. C. and R. A. acknowledges IISER-TVM for the financial support.
W.Z. acknowledges support from Research Starting Grants from South China Normal University (Grant No. 8S0340),
 the Natural Science
Foundation of Guangdong Province, China (Grant No. 2019A1515011868), and the National Natural Science Foundation of China (Grant No. 12075089).
V. K. C. thank DST, New Delhi for computational facilities under the DST-FIST programme (SR/FST/PS-1/2020/135) to the Department of Physics.
The work of V. K. C. is also supported by the SERB-DST-MATRICS Grant No.
MTR/2018/000676 and  CSIR Project under Grant No. 03(1444)/18/EMR-II.  DVS  is supported by the DST-SERB-CRG Project under Grant No. CRG/2021/000816.


%

\end{document}